# AI Literacy for Legal AI Systems: A practical approach

Gizem GÜLTEKIN-VÁRKONYI*

> "- I call the court to session….
> - The court is in session. What is the citizen robot's request?"
> - You are a Make Seven, designed in the Fourth Age.
> Your Legal Program would not contain the charge. Consult your archives."[1]

**Abstract**

Legal AI systems are increasingly being adopted by judicial and legal system deployers and providers worldwide to support a range of applications. While they offer potential benefits such as reducing bias, increasing efficiency, and improving accountability, they also pose significant risks, requiring a careful balance between opportunities, and legal and ethical development and deployment. AI literacy, as a legal requirement under the EU AI Act and a critical enabler of ethical AI for deployers and providers, could be a tool to achieve this. The article introduces the term "legal AI systems" and then analyzes the concept of AI literacy and the benefits and risks associated with these systems. This analysis is linked to a broader AI-L concept for organizations that deal with legal AI systems. The outcome of the article, a roadmap questionnaire as a practical tool for developers and providers to assess risks, benefits, and stakeholder concerns, could be useful in meeting societal and regulatory expectations for legal AI.

1. ## Introduction

AI systems are being adopted at an increasing rate in judicial and legal contexts worldwide. These systems aim to improve the efficiency and accuracy of legal and court procedures, as well as the decision-making processes involved. They are used for several administrative tasks, including case management, legal research, contract analysis, fraud detection, and document review[2] in overburdened legal entities such as law firms, courts, and corporate legal departments (hereafter: organizations). This article primarily focuses on AI systems that predict case outcomes with the aim of helping courts and law firms to assess cases primarily for decision-making purposes. Additionally, the article looks at other AI systems used or planned for use in the legal and judicial contexts, which could significantly affect involved parties, third parties, and the public, depending on the both benefits and risks aspects identified in this article, namely, bias, explainability, and time and cost evaluation. While the term "legal AI systems", as defined in this article, may be contested, its primary purpose is to ensure terminological consistency and facilitate comprehension. Moreover, the term "legal AI systems" could be used as an umbrella term for all AI systems that may cause bias, discrimination, opacity in explanations, and overreliance in judicial decision-making, as framed in this article. Besides the risks, legal AI systems also offer some benefits to providers and users, which could lead to a new world order in legal decision-making processes involving collective human-machine decision-making[3] and the article also take into consideration to improve this collaboration.

---

* assistant professor, University of Szeged, Faculty of Law and Political Sciences, International and Regional Studies Institute gizemgv@juris.u-szeged.hu
[1] Walter Tevis, *Mockingbird*. RosettaBooks LLC, 1980. ISBN: 978-07953430255.
[2] IBA and CAIDP, 'The future is now: Artificial intelligence and the legal profession', (2024). https://www.ibanet.org/document?id=The-future-is-now-AI-and-the-legal-profession-report.

The emergence of legal AI systems is beginning to affect how justice systems function and are organized. While legal AI systems have been introduced at the preliminary level in courts, they are expected to prompt organizational changes within entities that intend to use or develop them[4]. These changes are not optional, but rather legal obligations as stipulated by the European Union's Artificial Intelligence Act (hereafter: AI Act)[5]. This legislation is expected to have a profound impact on organizational processes by imposing a range of obligations on both providers and deployers of legal AI systems. Beyond administrative requirements, the AI Act also mandates changes in organizational practices and ethical standards. Among these obligations, Article 4 of the AI Act introduces the requirement of AI literacy (hereafter: AI-L) for organizations, underscoring the need for adequate knowledge and competence in the development and use of AI technologies. AI-L obligates organizations intending to provide and/or deploy legal AI systems to evaluate the risks and benefits associated with the tools, assess their staff's competencies, and develop an educational curriculum to enhance their AI-L competencies. As the explanation of the AI-L foresees contextuality in the AI Act, field specific researches is crucial. Existing literature does not provide a roadmap for organizations that intend to implement legal AI systems, and the AI Office which is tasked with supervision and enforcement of the Act, would not impose field-specific requirements for compliance based on the reasons that the AI-L is contextual[6]. Furthermore, there is no practice found in the recently published living repository on any type of legal AI systems[7]. Consequently, the present article aims to address this lacuna by introducing the concept of AI-L, the concept of legal AI systems and elucidating their relationship within organizations from specific risks and benefits dimensions, particularly, bias, time and cost efficiency, explainability dimensions. Some of these risks stem from the unpredictability of AI systems[8], particularly in case of large language models (LLMs), which acquire new data and use output data for training. This complicates the explanation of the entire processing activity, the detection of anomalies, and the immediate intervention required. Other risks arise from nontechnical issues, such as overreliance on these systems and by based on the assumption that they are time and cost efficient for the organizations. After assessing the benefits and risks, this article offers a set of questions that could be integrated into an organization's AI-L as a starting point.

---

[3] Anna Beckers and Gunther Teubner, 'Human–algorithm hybrids as (quasi-) organizations? On the accountability of digital collective actors', 50(1), *Journal of Law and Society,* (2024), 100–119. https://doi.org/10.1111/jols.12412

[4] European Commission, 'AI Literacy - Questions & Answers', 2025. https://digital-strategy.ec.europa.eu/en/faqs/ai-literacy-questions-answers

[5] Commission Regulation (EU) 2024/1689 of the European Parliament and of the Council of 13 June 2024 laying down harmonised rules on artificial intelligence and amending Regulations (EC) No 300/2008, (EU) No 167/2013, (EU) No 168/2013, (EU) 2018/858, (EU) 2018/1139 and (EU) 2019/2144 and Directives 2014/90/EU, (EU) 2016/797 and (EU) 2020/1828 PE/24/2024/REV/1. OJ L, 2024/1689, 12.7.2024.

[6] European Commission (2025)

[7] European Commission, 'Living repository to foster learning and exchange on AI literacy', 4 February 2025. https://tinyurl.com/2hah9evj

The article's limitation is that it does not aim to propose the implementation of full-speed AI systems in justice without acknowledging the inherent risks associated with entrusting judicial, administrative, and prosecutor functions to technological systems. Instead, it advocates for a more nuanced approach, emphasizing the role of technology as an enabler, not a substitute, for human professionals in judicial-related tasks. From this perspective, judicial automation initiatives must be designed from a pluralistic standpoint that acknowledges the diversity of legal traditions[9], and at this point, AI is not a voice to be considered.

## 2. Legal AI systems: A definition

The integration of AI into the legal and judicial sectırs has given rise to various legal AI systems, each designed to fulfill specific functions within judicial contexts. Several scholars have attempted to categorize these systems to understand and regulate their growing influence on legal services and adjudication. One of the earliest attempts proposes a functional typology based on the role of AI in digital legal services[10]. This classification ranges from basic digital legal tools that do not involve algorithmic processing, such as some online dispute resolution platforms, to advanced systems that generate algorithmic recommendations on judicial outcomes or operate as autonomous online courts[11]. In addition to the aforementioned service-oriented classification, a task-based typology has been proposed[12]. This typology focuses on the outcomes designed to be facilitated by AI systems. This approach encompasses three primary types of legal analysis: identification, categorization, and forecasting. Identification is defined as the process of recognizing key elements or conclusions in legal texts. This may include synthesizing case law or identifying legally relevant arguments. Categorization is defined as the process of analyzing the factors that influence legal decisions. These factors include the facts of the case, the underlying legal reasoning, and the argument's structure.

---

[8] Jason Millar and Ian Kerr, 'Delegation, relinquishment, and responsibility: The prospect of expert robots', in Ryan Calo, A. Michael Froomkin, and Ian Kerr (eds.), *Robot Law*, Cheltenham, UK: Edward Elgar Publishing (2016). https://doi.org/10.4337/9781783476732.00012
[9] Katalin Kelemen and Luis de Miranda, 'Courts as anthrobots: Learning from human forms of interaction to develop a philosophically healthy model for judicial automation', *15*(2), *International Journal for Court Administration,* (2024), 2. https://doi.org/10.36745/ijca.525
[10] Sourdin (2021)
[11] Ibid.
[12] Masha Medvedeva, Martijn Wieling and Michel Vols, 'Rethinking the field of automatic prediction of court decision', 31, *Artificial Intelligence and Law,* (2023), 195-212. https://doi.org/10.1007/s10506-021-09306-3

Forecasting uses predictive analytics to estimate probable legal case outcomes and facilitates strategic decision-making for those involved in legal proceedings.[13]

An example of integrating all these functionalities can be seen in China's so called "Smart Courts."[14] Since adopting these systems nearly a decade ago[15], the courts have provided virtual judicial services via AI for case law searches, document processing, and decision predictions[16]. In Europe, some member states have been actively piloting and deploying various automated and semi-automated tools, and AI systems for purposes including legal information retrieval, legal analytics, and speech-to-text conversion[17]. Some of these tools serve as decision support systems, while others are embedded in broader digital judicial infrastructures. In Germany, several companies are developing software that can autonomously analyze judgments and formulate statements about the future based on previous judgments[18]. France is rapidly advancing the integration of AI systems within the legal context, with private companies providing tools for contract management, predictive justice, legal strategy, and decision-making through generative AI technology and data-driven insights. [19]

The classification of legal AI systems remains a significant challenge due to their diverse technical configurations and functionalities. Some systems rely on rules-based algorithms and therefore may not fully meet formal definitions of AI under current regulatory frameworks. Despite this, such tools, when employed by judicial actors, can still have substantial effects on litigants, third parties, and the public. The increasing integration of LLMs, such as GPT, into legal and judicial contexts further complicates the task of establishing consistent terminology[20]. A number of legal and judicial actors have adopted LLMs under the broader umbrella of data analysis, employing them in tasks ranging from contract drafting to litigation strategy development to case prediction purposes[21]. The accessibility of these tools to all legal professionals, coupled with the potential for deceptive novel outputs, raises concerns about their potential adverse effects. [22] Although some legal professionals regard these tools as useful for certain tasks such as for legal interpretation[23], there are documented cases where their application has led to negative consequences for both lawyers and judges. [24] This widespread and varied use of GPT-based technologies has intensified the difficulty of classifying such tools within a unified conceptual framework. While terms like hybrid AI[25] have been proposed to capture this complexity, no consensus has yet emerged. In addition, the open accessibility of these models, combined with their potential to generate misleading outputs, raises ethical and legal concerns. All of the concerns mentioned should be evaluated and integrated into the professional context through education, which could enable professionals to identify and intervene in a timely manner. AI-L could be a good opportunity for this.

---

[13] Ibid
[14] Changqing Shi et al., 'The smart court – A new pathway to justice in China?' 12(1), 4, *International Journal for Court Administration*, https://doi.org/10.36745/ijca.367
[15] Yadong Cui, '*Artificial intelligence and judicial modernization*', Springer Nature Singapore Pte Ltd, (2020). https://link.springer.com/book/10.1007/978-981-32-9880-4
[16] Tara Vasdani, 'Robot justice: China's use of Internet courts', *The Lawyer's Daily,* 5 February 2020. https://www.thelawyersdaily.ca/articles/17741/robot-justice-china-s-use-of-internet-courts
[17] CEPEJ, 'Resource Centre Cyberjustice and AI'. https://tinyurl.com/mr3h8j8h
[18] International Bar Association guidelines and regulations to provide insights on public policies to ensure artificial intelligence's beneficial use as a professional tool. https://www.ibanet.org/PPID/Constituent/Multi-displry_Pract/anlbs-ai-report
[19] Ibid.
[20] Jinqi Lai et al., 'Large language models in law: A survey', (5). *AI Open,* (2024), 181–196. https://doi.org/10.1016/j.aiopen.2024.09.002
[21] Tom Saunders, 'Legal tech teams turn to AI to advance business goals', *Financial Times*, 19 October 2023. https://www.ft.com/content/9a117ac7-29ae-43fe-b840-a04005b98799
[22] Mathieu Veron, 'Do this LLMs use my prompt data for training' *Medium,* 2 July 2024. https://tinyurl.com/587346s3
[23] Nate Raymond, 'US judge runs 'mini-experiment' with AI to help decide case', *Reuters*, 6 September 2024. https://tinyurl.com/pfjr6aab
[24] Hibaq Farah, 'Court of appeal judge praises 'jolly useful' ChatGPT after asking it for legal summary', *The Guardian*, 15 September 2023. https://tinyurl.com/bdfd7jm8

Building on the evolution and typologies of legal AI systems presented, the article further focuses on their placement within the AI Act, particularly from the perspective of AI-L. Regardless of their risk classification, most of the legal AI systems will probably be required to comply with the AI-L obligations set forth in the Act and further specified by the European Commission[26]. To fully grasp the implications of this requirement, it is essential to examine the role and categorization of legal AI systems within the broader framework of the AI Act.

### 2.2. Legal AI Systems in the AI Act

It is a risk not to be aware of the potential risks and benefits of a particular AI system. The AI Act adopts an exact risk-based approach, classifying AI systems according to their potential impact. The objective of this approach is to identify and manage risks prior to market launch of an AI tool, thereby facilitating the responsible use of AI systems throughout the European Union. AI systems, in the AI Act, are categorized into four levels of risk: unacceptable risk, high risk, limited risk, and minimal risk. The organizations and persons involved in the life-cycle of AI systems falling under one of these categories are required to fulfill proportionate obligations introduced in the AI Act. These obligations, at glance, aim to ensure safety, accountability, and respect for fundamental rights.

Legal AI systems, firstly, are likely to fall within the category of high-risk AI systems under the AI Act, as their outputs may affect individuals' fundamental rights**.** Annex III, paragraph 8(a), specifically lists AI systems used in the administration of justice and democratic processes as high-risk, as referenced in Article 6(2) of the Act. According to Recital 61, while non-binding, such systems are considered high-risk due to their potential to introduce bias, error, and opacity into judicial decision-making. This classification is particularly relevant where AI is used by judicial authorities, or on their behalf, to support the interpretation of legal principles**,** fact-finding**,** or the application of law to individual cases. It also extends to the use of AI in dispute resolution contexts. For instance, legal AI systems employed for predictive decision support in criminal courts clearly fall within this high-risk category, given their significant implications for liberty, fairness, and procedural rights.

---

Molly Bohannon, 'Lawyer Used ChatGPT In Court And Cited Fake Cases: A Judge Is Considering Sanctions', *Forbes,* 8 June 2023. https://tinyurl.com/5axdnd8n

[25] Fotis Fitsilis and George Mikros, 'AI-based solutions for legislative drafting in the EU Summary report' European Commission Directorate-General for Digital Services, (2024). https://tinyurl.com/3y7ztkkk

[26] European Commission (2025)

Secondly, despite the fact that certain legal AI systems may not fall directly into the high-risk category, it is imperative to implement AI-L for all categories.[27] AI-L is a rule that applies to all categories, as the wording of Article 4 does not refer to any risk categories. Further, Article 95 reading together with Recital 165 of the AI Act foresees that AI-L is applicable to both non-high-risk categories as well, albeit in an obligatory manner. This would be a logical approach since the high-risk status of a system today may not persist into the future is salient and the risk categories delineated by the AI Act do not present a clear methodology for their identification[28]. Ultimately, the objective of AI-L is to facilitate risk and benefit assessments, thereby ensuring the comprehensibility of the system and establishing the reliability of the system.

For the organizations, a lack of awareness regarding the risks associated with legal AI systems may result in a range of consequences. Failure to fulfill legal obligations, financial losses, or diminished competitiveness due to missed opportunities are all possible consequences which lead loss of trust in the organization. [29] Judicial and legal organizations are among the entities that must be entrusted with the highest level of confidence in a democratic society. AI-L serves as a tool for the organizations to identify their awareness of this situation. As delineated in Article 4 of the AI Act, which will be expounded upon subsequently, the EU law-maker has devised a novel approach to ensure that organizations adhere not only to legal requirements but also to educate its staff and other persons dealing with the operation and use of AI systems on their staff. From this perspective, the AI-L, aims of enabling them to comprehend, assess, and navigate the benefits, risks, and limitations of AI systems. This competence is especially crucial for legal AI deployers and providers, as their systems must meet high-risk criteria that demand specialized knowledge of compliance, ethics, and risk mitigation.  In a manner similar to the GDPR[30], which established the role of data protection officers within organizations (Article 37) and mandated Data Protection Impact Assessments (Article 35), introducing a novel organizational framework, the EU AI Act likewise indicates a requirement for the implementation of robust, clearly delineated roles for AI.

3. **Concept of the AI Literacy in the AI Act**

The term AI-L has already been referred to by some international organizations before, such as UNESCO[31] characterizing AI-L as being interrelated with the knowledge, skills, and values necessary to responsibly use, evaluate, and engage with AI systems. The World Economic

---

[27] ibid.
[28] Martin Ebers, 'Truly Risk-Based Regulation of Artificial Intelligence How to Implement the EU's AI Act', *European Journal of Risk Regulation*, (2024), 1–20. https://doi.org/10.1017/err.2024.78.
[29] Decoding responsibility in the era of automated decisions: Understanding the implications of the CJEU's SCHUFA judgment. Centre for Information Policy Leadership, 2024. https://tinyurl.com/4van5x3x
[30] Regulation (EU) 2016/679 of the European Parliament and of the Council of 27 April 2016 on the protection of natural persons with regard to the processing of personal data and on the free movement of such data, and repealing Directive 95/46/EC (General Data Protection Regulation), OJ L 119, 04/05/2016, p. 1–88.
[31] Beijing consensus on artificial intelligence and education. UNESCO, 2019. https://unesdoc.unesco.org/ark:/48223/pf0000368303.

Forum[32] also articulated the universal necessity of AI-L, characterizing it as the capacity to comprehend and assess the advantages, risks, and constraints of AI through accessible, contextualized learning, with an emphasis on practical applications and inclusivity for marginalized groups. The AI Act, in its capacity as a legal concept, defines AI-L in Article 4 as:

> "skills, knowledge and understanding that allow providers, deployers and affected persons, taking into account their respective rights and obligations in the context of this Act, to make an informed deployment of AI systems, as well as to gain awareness about the opportunities and risks of AI and possible harm it can cause."

The definition comprises three primary components that are designed to equip stakeholders (i.e., deployers, providers, and affected persons) with the necessary skills, knowledge, and understanding of the system. First, it highlights the protection of the affected persons since they are indirectly protected by the provisions of the Act, and then further highlights obligations of the deployers and providers. Second, it emphasizes the objective of informed deployment of AI systems. Third, it aims to raise awareness of the risks, opportunities, and potential harm associated with AI. Furthermore, Recital 20 of the Act offers a concise yet ambiguous delineation of the concept, emphasizing the responsibilities of providers and deployers to ensure the AI-L by considering the technical knowledge, experience, education, and training of the individuals involved. This involves internal teams, ranging from technical teams to management directly handling or overseeing AI systems, which essentially concerns the organization as a whole. The measures should account for the AI-L refer to an effective communication within (e.g., provider) and in between (provider-deployer) organizations .The purpose of this paper is not to adopt a definition, since the paper works on the legal aspect which includes the fact that the term includes being aware of risks and opportunities. Further, guidelines and explanations from the official EU authorities, such as the EU AI Board or AI Office, or relevant national bodies, will be considered for practices. These guidelines are not yet available, despite the AI Act's Chapter I, which includes the AI-L rule, entered into force in February 2025. Since then, only a couple of practical outputs have been delivered. The EU AI Office held an online event in which practical examples from the field were shared[33] and an explanatory Q&A document from the Commission was published[34] (though the mode of choosing questions is unclear), besides previously mentioned living repository with the aim of collecting business practices (rather than explaining the concept) have recently been published. The absence of detailed guidelines from EU authorities creates challenges in planning the AI-L within the organizations, and later will be challenging implementation. In this case, the concept of AI-L in the EU AI Act might be considered as analogous to the AI-L in the literature. Definitions, concepts, and the close relationship between ethics and AI-L in the literature could be a suitable starting point.

---

[32] Without universal AI literacy, AI will fail us. World Economic Forum, 2022.
https://www.weforum.org/stories/2022/03/without-universal-ai-literacy-ai-will-fail-us/
[33] https://www.youtube.com/watch?v=Dyf4ZVts9HY
[34] European Commission (2025)

## 3.1. AI Literacy in the Literature

A common approach in the literature to define AI-L involves conducting systematic literature reviews that analyze existing definitions and approaches. These reviews identify competencies and factors relevant to AI-L and often develop scales to measure AI-L levels, sometimes applying them in specific domains, and they typically distinguish between experts and non-experts in AI. A close examination of the extant literature reveals a persistent nexus between AI-L and the identification of risks and opportunities associated with AI in organizational settings.

The scope of the initial literature on AI-L explores the term within educational settings, examining how it can be defined and tailored to different levels, such as k-12,k-16, higher education, and even law schools[35]. These studies investigate why AI-L is necessary for specific educational contexts and how it can be designed to meet their needs. More recent and broader approaches emphasize defining AI-L through competencies that enable individuals to critically evaluate AI technologies from the risks and benefits point of view, communicate and collaborate effectively with AI, and use AI as a tool across various domains, such as workplaces and homes. For instance, an often cited work[36] in literature defines AI-L as a set of competencies that dispel misconceptions, promote critical use of AI, and foster a more diverse AI workforce. Their work also establishes a connection between AI-L and organizational readiness for the AI era, emphasizing the significance of capabilities such as understanding existed considerations, human-machine interaction, and dynamic learning strategies. Taking a similar approach to that, the discussions drawn in this article are of particular importance for organizations integrating AI systems, as they ensure employees can engage with both the advantages and disadvantages of these technologies.

In the context of the AI-L competencies, another frequently cited work[37] identifies essential components of AI-L as it is a tool for cultivating responsible citizens capable of using AI in trustworthy and fair ways. While the authors initially referenced this framework within the context of education and expertise, its relevance could be expanded to encompass a broader range of organizations, including legal AI providers and deployers. In this regard, AI-L functions as a capacity-building instrument, aimed at fostering recognition, utilization, and evaluation of AI technologies in a manner consistent with the imperative to comprehend the risks and responsibilities associated with AI. [38] Another work provides a collection of AI-L

---

[35] Lorena Casal-Otero et al., 'AI literacy in K-12: A systematic literature review', *International Journal of STEM Education, 10*, (2023), 29. https://doi.org/10.1186/s40594-023-00418-7

Davy Tsz Kit Ng et al., 'Conceptualizing AI literacy: An exploratory review', 2, *Computers and Education: Artificial Intelligence*, (2021),100041. https://doi.org/10.1016/j.caeai.2021.100041

Michal Černý, 'AI literacy in higher education: Theory and design'. In Ł. Tomczyk (Ed.), *New media pedagogy: Research trends, methodological challenges, and successful implementations. Communications in Computer and Information Science,* (2024), *2130*. Springer. https://doi.org/10.1007/978-3-031-63235-8_24

Sara Migliorini and João Ilhão Moreira, J. I., 'The case for nurturing AI literacy in law schools', *11*(1), *Asian Journal of Legal Education,* (2024). https://doi.org/10.1177/23220058241265613

[36] Duri Long and Brian Magerko, 'What is AI literacy? Competencies and design considerations', In *Proceedings of the 2020 CHI Conference on Human Factors in Computing Systems*, New York, NY: ACM, (2020). https://doi.org/10.1145/3313831.3376727

[37] Ng et al. (2021)

[38] Bingcheng Wang, Pei-Luen Patrick and Tianyi Yuan, 'Measuring user competence in using artificial intelligence: Validity and reliability of artificial intelligence literacy scale', *42*(9), *Behaviour & Information Technology,* (2022). https://doi.org/10.1080/0144929X.2022.2072768

definitions for experts and non-experts, incorporating ethics and law.[39] AI-L's role in enhancing explainability and transparency in the AI systems, together with a reference to organizations responsibility to switch their focus more on these concepts has also been noted[40]. This conceptualization reflects that AI-L is expected to impact organizations by requiring them to implement numerous procedures to apply in terms of complying with AI-L requirements.

### 3.2. Evaluating Legal AI Systems through AI-L at Organizations

Academic literature directly examining the organizational effects of AI-L remains is limited, however, growing attention is expected with the AI Act's formal obligation for organizations to foster it. Early approaches frame AI-L as essential to how institutions adapt to the AI era, emphasizing the acquisition of competencies that enable critical evaluation of AI's capabilities and limitations[41]. For example, human-machine interaction and learning capabilities help employees understand algorithmic implications, while principles such as transparency and explainability[42] are vital for organizations implementing or designing legal AI systems. Without integrating such strategies, organizations risk falling behind in an increasingly digital legal environment[43] besides failing to comply with legal or ethical principles.

AI-L aligns with key tenets of responsible AI which are fairness, accountability, and social justice.[44] It is crucial for non-technical actors, such as lawyers and judges, to be able assess algorithmic outcomes, understand bias and representation issues, and anticipate long-term impacts to ensure responsible AI principle. A stakeholder-first approach encourages organizations to address the informational needs of workers, policymakers, and the public, thereby enhancing ethical engagement and trust[45]. Operationalizing AI-L involves addressing both technical and non-technical staff's informational needs.[46] Several studies propose measurement scales targeting non-experts' ability to evaluate AI-generated decisions[47], especially in high-impact fields like generative AI[48]. These tools also address legal comprehension, regulatory compliance, and risk awareness. Domain-specific AI-L, particularly in the legal field, is best assessed through integrated tools that combine general, sectoral, and ethical dimensions[49]. This recommends diverse and refined assessment methods to assess staff's AI-L literacy level with tool such as multiple-choice tests, behavioral evaluations, and self-assessments to capture both theoretical knowledge and applied competence. Other research suggests AI-L should be viewed as a blend of technological understanding, evaluative skill, and application-specific knowledge, enabling staff to select appropriate systems and assess their organizational impact for a broader societal impact[50]. All of the presented concepts reflect the idea that the implementation of AI-L cannot be limited to a single definition, practice, or context. The multi-contextual nature of AI systems requires engagement with different stakeholders to identify an AI-L framework in context-specific.

---

[39] Marc Pinski and Alexander Benlian, 'AI literacy for users – A comprehensive review and future research directions of learning methods, components, and effects', 2(1), *Computers in Human Behavior: Artificial Humans,* (2024), 100062. https://doi.org/10.1016/j.chbah.2024.100062

[40] Andrew Cox, 'Algorithmic literacy, AI literacy, and responsible generative AI literacy', 18(3), *Journal of Web Librarianship,* (2024), 98. https://doi.org/10.1080/19322909.2024.2395341

[41] Long and Magerko (2020)

[42] Ibid. 9.

[43] Dilek Çetindamar et al., 'Explicating AI literacy of employees at digital workplaces', 71(3), *IEEE Transactions on Engineering Management,* (2024), 810–823. https://doi.org/10.1109/TEM.2021.3138503

[44] Daniel Domínguez Figaredo and Julia Stoyanovich, 'Responsible AI literacy: A stakeholder-first approach', 10(2), *Big Data & Society,* (2023).. https://doi.org/10.1177/20539517231219958

[45] Ibid.

[46] Celalettin Çelebi et al., 'Artificial intelligence literacy: An adaptation study', 4(2), *Instructional Technology and Lifelong Learning*, (2023), 291–306. https://doi.org/10.52911/itall.1401740
Wang et al. (2022)

[47] Matthias Carl Laupichler et al., 'Development of the "Scale for the assessment of non-experts' AI literacy" – An exploratory factor analysis', 12, *Computers in Human Behavior Reports,* (2023), 100338. https://doi.org/10.1016/j.chbr.2023.100338

[48] Ravinithesh Annapureddy, Alessandro Fornaroli and Daniel Gatica-Perez, 'Generative AI literacy: Twelve defining competencies', 6(1), *Digital Government: Research and Practice*, (2024). https://doi.org/10.1145/3685680

In the following step, all foundational aspects reviewed thus far will be integrated through an assessment of the interplay between the risks and benefits of legal AI systems. This assessment will focus on the following components: explainability, time and cost efficiency, and the mitigation of bias and discrimination. These components are selected because they represent key indicators of both ethical robustness and practical effectiveness in the deployment of AI within legal settings. The objective is to provide a nuanced perspective on optimizing legal AI systems in a manner consistent with the principles of transparency, fairness, and accountability at the organizational level, using AI-L components as a guiding framework.

## 4. Legal AI Systems Benefit Assessment

The aim of this part is to present the benefits that organizations operating in the legal or judicial field could gain when they use legal AI systems. Since AI-L is not only about risk assessment, but also discovering the benefits of these tools, benefit assessment which is presented under this title could be one component of AI-L for the organizations.

### 4.1. Benefit I: Eliminating Bias and Discrimination

If there is one fate that humans and algorithms share, it is bias and discrimination. As long as there is human bias, there will be bias in algorithmic outputs as well. In 1931, in the United States, out of nine African-American teenagers, eight were sentenced to death and one to life imprisonment, falsely accused of raping two white women on a train from Tennessee to Alabama. As it later turned out, clear procedural violations and a lack of evidence that their skin color created in the eyes of not only the claimants but also the judge led to this decision, even though this verdict was initially overturned by the Supreme Court. Also known as the *Scottsboro Boys*[51], the case created many precedents for resolving racism and right to fair trial in the judicial process in the US.

Human judges, even they are professionals, are not fundamentally different from other individuals when it comes to analyzing human behavior and making decisions[52]. They are influenced by many factors such as personal, societal, and cultural predispositions when making a decision.[53] In addition to physical and physiological factors such as heat or neural fatigue caused by heavy caseloads, the impact of attributes such as gender and ethnicity where the judge comes from can lead a judge to make some biased decisions.[54] For example, in some countries where the democratic context is weak, wealth, appearance, or social status may unduly influence judicial outcomes, allowing for manipulation and undermining fairness. In another example, imagine a law enforcement agency asking about people with certain religious backgrounds to an LLM that fulfills the requests by generating violent scenarios, including bombings and attacks, as they are the most relevant to the training data coming from the society collected.[55]

---

[49] Nils Knoth et al., 'Developing a holistic AI literacy assessment matrix: Bridging generic, domain-specific, and ethical competencies', 6, *Computers and Education Open,* (2024), 100177.
https://doi.org/10.1016/j.caeo.2024.100177

[50] Ismail Çelik, 'Exploring the determinants of artificial intelligence (AI) literacy: Digital divide, computational thinking, cognitive absorption', 83, *Telematics and Informatics,* (2023), 102026.
https://doi.org/10.1016/j.tele.2023.102026

[51] *Powell v. Alabama,* 287 U.S. 45 (1932)

[52] Katherine B. Forrest, '*When machines can be judge, jury, and executioner: Justice in the age of artificial intelligence',* World Scientific Publishing Company, (2021) ISBN: 9789811232725.

Unlike humans, algorithms are immune to stress, hunger, and emotional disturbance, and their errors, while possible, can often be systematically addressed through monitoring and refinement. When developed using accurate, representative data and free of demographic markers such as religion, ethnicity, or gender, they can make decisions unaffected by fatigue, emotion, or external pressures. An analysis[56] of biases and prejudices reflected by juries, whose output affects the court's ultimate decision, could be partially overcome by AI systems. Re and Solow-Niederman (2021)[57] argue that encoding standardized justice systems into algorithms can improve consistency and fairness, provided that these systems are regularly tested to prevent evolving biases. Volokh (2019)[58] similarly advocates for regular scenario-based testing to ensure objectivity, an ideal that is often unattainable for human judges. If the training data is free of bias and discrimination, and if it is sustainable, legal AI systems could open a new chapter in the protection of human rights and fair procedures and outcomes.

## 4.2. Benefit II: Time and Cost Efficiency

Justice is not for everyone and it is not free when it comes to calculating the cost and length of legal proceedings. Depending on the case, it can be very costly and time consuming for an individual to find ways to defend the rights granted them by the law. Due to the high cost of legal advice, only the wealthy have effective access to justice, while the poor rely on legal aid in court cases[59]. The cost of basic legal advice per hour, the type of official document to be drafted, and other costs of conducting a legal proceeding, as determined by a country's bar association, could provide an indication of how much an individual would be charged for even a basic legal proceeding, which could sometimes exceed a minimum wage. It costs also for organizations to provide legal services, but AI has the potential to reduce costs and increase revenues for companies that offer legal services or have legal service departments.[60] Legal AI systems can also offer time and cost efficient legal services for the organizations and individuals.

---

[53] Brian M. Barry, '*How judges judge: Empirical insights into judicial decision-making*', Routledge, (2021), ISBN: 9780429023422.
[54] Ibid.
[55] Andrew Myers, 'Rooting out anti-Muslim bias in popular language model GPT-3', *Stanford HAI,* 3 July 2022. https://hai.stanford.edu/news/rooting-out-anti-muslim-bias-popular-language-model-gpt-3
[56] Matthew J. O'Hara, 'I, for one, welcome our new AI jurors: ChatGPT and the future of the jury system in American law', 3(2), *International Journal of Law, Ethics, and Technology,* (2024), 50-84. https://www.ijlet.org/4.3.2
[57] Richard M. Re and Alicia Solow-Niederman, 'Developing artificially intelligent justice', 22(2), *Stanford Technology Law Review,* (2019), 242–289. https://www.ijlet.org/wp-content/uploads/2025/01/IJLET-4.3.2.pdf
[58] Eugene Volokh, 'Chief justice robots', 68 (6), *Duke Law Journal*, (2019), 1135–1192.
[59] Gravett, W. H. (2020). Is the dawn of the robot lawyer upon us? The fourth industrial revolution and the future of lawyers. Potchefstroom Electronic Law Journal, 23, 1–37. https://scholarship.law.duke.edu/dlj/vol68/iss6/2
[60] 2024 in charts. McKinsey Global Institute, 2024. https://www.mckinsey.com/mgi/our-research/mckinsey-global-institute-2024-in-charts

The length of legal proceedings can be attributed to a number of factors, including administrative inefficiencies, staff shortages, delays in gathering evidence, and the complexity of the case. Complex legal matters, particularly in international and criminal law, often require extensive research, which can be time-consuming and costly, whether conducted by the lawyer or delegated to colleagues or interns. Violation of Article 6 of the ECHR, which concerns the length of proceedings by states, is one of the most common complaints submitted to the ECHR in 2023.[61] Combined with unethical practices in the legal world, e.g. in commercial disputes, lawyers who charge hourly fees might benefit from prolonging conflicts for financial gain[62], justice might be heard as the ugly face of the legal community and losing time.

Algorithms could overcome the cost and length for legal services in several ways. For example, there are some initiatives and work showing that electronic case management and modeling of court rules could increase efficiency[63] in terms of speed of delivery of legal services. The DoNotPay platform is a prime example of the expanding role of AI in legal services. Originally designed to mediate disputes over parking tickets, the system now addresses a range of issues, including assistance in applying for government benefits and even human rights complaints. In addition, LLMs, for example, help overcome language barriers in multilingual jurisdictions, providing faster access to justice, [64]and assist in providing faster support for legal services to citizens.[65]

For humans, it takes time to get familiar with a new legislation, analyze its implications, and apply it in practice[66]. In contrast, AI can process the same information in minutes or, at most, a few hours. AI-powered legal tools can significantly streamline the process by assisting lawyers in gathering information, categorizing cases or analyzing precedents, and predicting case outcomes. In Anglo-Saxon law, for instance, AI can efficiently handle precedent-based reasoning since the cases are linked to each other. An AI system was developed[67] on the decisions of the Federal Supreme Court demonstrated its ability to access the decisions of the judges who made the unquestionably correct decisions with 72% accuracy (the question of whether the remaining 18% gap is acceptable remains a matter of debate).

Finally, saving time and money in court proceedings and legal services would, on the one hand, contribute to justice and democracy, as legal AI systems could improve fair access to justice and, in some cases, democracy.[68] On the other hand, judges and lawyers may be able to separate more time for practicing their profession from the perspective of contributing to law and justice. As Markovic argues[69] 'creativity increases in areas where automation is introduced' and if this creativity is applied to law and justice, it could enhance the alignment of legislation with its intended purpose through the active involvement of legal professionals in lawmaking. Furthermore, Winter (2021)[70] predicts that the adoption of AI-powered justice systems, including robotic judges, could reduce judicial caseloads, allowing lawmakers and legal actors to focus on designing better laws rather than managing minor disputes.

---

[61] Annual report 2023, ECtHR. https://www.echr.coe.int/documents/d/echr/annual-report-2023-eng
[62] Richard Susskind, *'Online Courts and the Future of Justice'*, New York, Oxford Academic, (2019) online edn. https://doi.org/10.1093/oso/9780198838364.001.0001
[63] Joshua C. Fjelstul, Matthew Gabel and Clifford J. Carrubba, 'The timely administration of justice: Using computational simulations to evaluate institutional reforms at the CJEU', 30 (12), *Journal of European Public Policy,* (2022), 2643–2664. https://doi.org/10.1080/13501763.2022.2113115
Caio Castelliano, Peter Grajzl, and Eduardo Watanabe, 'Does electronic case-processing enhance court efficacy? New quantitative evidence', 40 (4), *Government Information Quarterly,* (2023), 101861. https://doi.org/10.1016/j.giq.2023.101861
[64] Abhinav Joshi et al. , 'IL-TUR: Benchmark for Indian legal text understanding and reasoning', *arXiv Preprint,* (2024). https://arxiv.org/abs/2407.05399
[65] Quinten Steenhuis and Hannes Westermann, 'Getting in the door: Streamlining intake in civil legal services with large language models', *arXiv Preprint*, (2024). arxiv.org/abs/2410.03762
[66] Susskind (2019).
[67] Daniel Martin Katz, Michael J. Bommarito and Josh Blackman, 'A general approach for predicting the behavior of the Supreme Court of the United States', 12 (4), *PLOS ONE* (2017). https://doi.org/10.1371/journal.pone.0174698

Of course, for all of this to happen, the systems need to be fed with clear and sufficient data, which is not currently the case, as not all information is accessible for reasons such as privacy, and the systems should not be hallucinated, which will be discussed later.

### 4.3. Benefit III: Improving Explainability

The principle of explainability is fundamental in legal systems and guarantees the individual's right to a reasoned decision. Rooted in the broader right to a fair trial, these principles guarantee access to and understanding of the reasoning behind judicial decisions. Courts around the world have embraced this principle by publishing decisions online, promoting democratic transparency and individual empowerment. Similarly, the duty of lawyers to inform their clients is consistent with this ethos of accessibility. However, it is not always easy for humans to explain their decisions[71], and this challenge could be partially overcome by AI. AI, if developed with relevant tools that improve explainability, has the potential to improve transparency as well.

Although AI systems have been referred to as black boxes from both technical and societal perspectives[72] because they operate in a way that obscures their decision-making processes, advances in AI design can counteract this opacity by employing techniques such as explainable machine learning, which produces results that are more accessible to users. From explainable AI initiatives to new original solutions for improving explainability in AI systems, there are both several approaches focusing on technical and social solutions to enhance explainability[73], including those for mitigating bias in LLMs.[74] A human-centered approach, additionally, is essential for AI design for achieving meaningful transparency.[75] Explainable AI systems and techniques could be used in legal AI by adopting a human-centered model that can increase understanding and trust in AI models. This could be done in many ways, for example, by revealing the importance of individual characteristics that are central to the decision and specific to the human being evaluated.[76] For this to happen, coordinated organizational action should be taken to ensure a human-centered approach. In the words of Schneiderman (2020)[77], 'human-centered AI begins when system designers, developers, and managers decide to involve a range of stakeholders in system design to create user-centered systems'.

---

[68] Brandon Long and Amitabha Palmer, 'AI and access to justice: How AI legal advisors can reduce economic and shame-based barriers to justice', 33(1), *TATuP - Zeitschrift für Technikfolgenabschätzung in Theorie Und Praxis*, (2024), 21–27. https://doi.org/10.14512/tatup.33.1.21

[69] Milan Markovic, 'Rise of the robot lawyers', 61 (2), *Arizona Law Review*, (2019), 325–350. https://arizonalawreview.org/pdf/61-2/61arizlrev325.pdf

[70] Christoph K. Winter, 'The challenges of artificial judicial decision-making for liberal democracy', In P. Bystranowski, P. Janik, & M. Próchnicki (Eds.), *Judicial decision-making: Integrating empirical and theoretical perspectives,* Springer Cham. vol.14., (2021). doi.org/10.1007/978-3-031-11744-2_9

[71] Volokh (2019).

[72] Andreas Matthias, 'The responsibility gap: Ascribing responsibility for the actions of learning automata', 6(3), *Ethics and Information Technology*, (2004), 175–183. https://doi.org/10.1007/s10676-004-3422-1
Aimee van Wynsberghe, 'Artificial intelligence: From ethics to policy', European Parliament Panel for the Future of Science and Technology, (2020). https://tinyurl.com/macpukkz
Vladislav V. Fomin and Paulius Astromskis, 'The Black Box Problem', in John-Stewart Gordon (ed.), *Future Law, Ethics, and Smart Technologies*. Leiden, The Netherlands: Brill, (2023). https://doi.org/10.1163/9789004682900_012

[73] Chris Reed, Keri Grieman and Joseph Early, 'Non-Asimov explanations: Regulating AI through transparency', 1, *The Swedish Law and Informatics Research Institute*, (2022), 315–338. https://doi.org/10.53292/208f5901.20b0a4e7
Shakti Kinger and Vrushali Kulkarni, 'Demystifying the black box: An overview of explainability methods in machine learning', 46 (2), *International Journal of Computers and Applications,* (2023), 90–100. https://doi.org/10.1080/1206212X.2023.2285533

In the legal domain, there are approaches to mitigate the black-box problem with strategies that identify commonly cited problems such as interpretability and unpredictability[78]. Legal AI systems, mostly, can solve this problem by providing personalized explanations that adapt to the user's level of expertise and address specific concerns. By involving judges, lawyers, and citizens in the design process, these systems can ensure that outputs reflect not only technical accuracy but also societal values, if they could be coded. Explainability tools such as user-friendly interfaces, visualizations, and natural language explanations can bridge the gap between complex algorithms and human understanding. These innovations are in line with the Explainable AI movement[79] and are intended to help build trust[80] and promote ethical use at organizations. As it is clear from all these statements, AI-L is an integral part of improving explanation, transparency, fairness, and accountability, as it requires deployers to understand the potential and consequences of the system.

## 5. Legal AI Systems Risk Assessment

The previous section aimed to show that if legal AI systems have a potential to ensure explainability at some level, while avoiding the discriminatory tendencies inherent in human decision making in a time and cost efficient manner. Here the discussion focuses on the challenges, both technical and social, and of practicing with legal AI systems at organizations. In the end, it should be clear to the reader that there are more risks that are difficult to mitigate compared to the benefits, and these risks need to be included in the organization's AI-L program.

---

[74] Mustafa Bozdağ, Nurullah Sevim and Aykut Koç, 'Measuring and mitigating gender bias in legal contextualized language models', 18(4), *ACM Transactions on Knowledge Discovery from Data,* (2024). https://doi.org/10.1145/3628602

[75] Jonas Rebstadt et al., 'Towards personalized explanations for AI systems: Designing a role model for explainable AI in auditing', 2, *Wirtschaftsinformatik Proceedings, (*2022). https://aisel.aisnet.org/wi2022/ai/ai/2

[76] Georgios Pavlidis, 'Unlocking the black box: Analysing the EU artificial intelligence act's framework for explainability in AI', 16(1), *Law, Innovation and Technology*, (2024), 293–308. https://doi.org/10.1080/17579961.2024.2313795

[77] Ben Shneiderman, 'Bridging the gap between ethics and practice: Guidelines for reliable, safe, and trustworthy human-centered AI systems', 10(4), *ACM Transactions on Interactive Intelligent Systems,* (2020). https://doi.org/10.1145/3419764

[78] Bartosz Brożek et al., 'The black box problem revisited: Real and imaginary challenges for automated legal decision-making', 32, *Artificial Intelligence and Law,* (2024), 427–440. https://doi.org/10.1007/s10506-023-09356-9

[79] Pantelis Linardatos,Vasilis Papastefanopoulos and Sotiris Kotsiantis,'Explainable AI: A review of machine learning interpretability methods', 23(1), *Entropy,* (2021), 18. https://doi.org/10.3390/e23010018

[80] Ronan Hamon et al., 'Bridging the gap between AI and explainability in the GDPR: Towards trustworthiness-by-design in automated decision-making', 17(1), *IEEE Computational Intelligence Magazine,* (2022), 72–85. https://doi.org/10.1109/MCI.2021.3129960

## 5.1. Risk I: Bias and Overreliance

Bias in human decision-making in the context of the acceptance of legal AI as an authority is one of the initial concerns. One such challenge stems from the perception that AI technology is infallible, which can lead to undesirable bias among individuals who use AI systems, as well as judges and lawyers who rely on AI assistance. For example, while recidivism prediction tools may offer remarkable efficiency,[81] their results may unduly influence decision makers. In *States v. Loomis* (881 N.W.2d 749 (Wis. 2016)), the COMPAS algorithm played a role in determining Mr. Loomis's risk of reoffending, raising concerns about potential violations of his right to a fair trial. Mr. Loomis argued that the algorithm relied on outdated data, discriminated based on gender, and generalized results based on aggregate data rather than personal assessments. Although higher courts upheld the use of the algorithm, a minority of judges expressed concern about its lack of transparency and the negative effect it made on judicial independence. Supporting the concerns raised in the Loomis case, recent studies support the idea of algorithmic bias in judicial decision-making[82], which compromises the principle of impartiality of the courts by adopting the algorithm as an authority as a result of its strong manipulative effect, a concept also known as 'obedience to the algorithm'[83]. This obedience is a consequence of many factors, but mostly has its roots in the early design of manipulative systems.

The anthropomorphic design of the algorithm is one of the many reasons why people unquestioningly obey in AI systems. The paradigm of Computers as Social Actors phenomenon[84] suggests that human-like attributes in machines foster misplaced trust, which could be true in today's AI systems, leading even experts into false and unconscious use with their manipulative generative responses[85]. La Rosa and Dank (2018)[86] further elaborated on this notion, asserting that trust in AI stems from the roles attributed to it and its perceived competence in fulfilling these roles. However, as Mori (1970)[87] theorized decades earlier, people's susceptibility to perceiving robots as human-like increases as these machines exhibit more anthropomorphic characteristics. In the context of legal AI systems, this misplaced trust could lead to unrealistic expectations, such as demands for empathy and compassion from systems rather than from judges and lawyers. All of these parameters could only contribute to public mistrust of organizations that provide and deploy legal AI systems.

---

[81] Zhiyuan Jerry Lin et al., 'The limits of human predictions of recidivism', 6(1), *Science Advances*, (2020). doi: 10.1126/sciadv.aaz0652
Fabian Lütz, 'The AI Act, gender equality, and non-discrimination: What role for the AI office?', 25, *ERA Forum,* (2024), 79–95. https://doi.org/10.1007/s12027-024-00785-w
[82] Manuel Portela et al., 'A comparative user study of human predictions in algorithm-supported recidivism risk assessment', *Artificial Intelligence and Law,* (2024). https://doi.org/10.1007/s10506-024-09393-y
[83] Maryam Ghasemaghaei and Nima Kordzadeh, 'Understanding how algorithmic injustice leads to making discriminatory decisions: An obedience to authority perspective', 61(2), *Information & Management,* (2024), 103921. https://doi.org/10.1016/j.im.2024.103921
[84] Clifford Nass and Youngme Moon, 'Machines and mindlessness: Social responses to computers', 56(1), *Journal of Social Issues*, (2000), 81–103. https://doi.org/10.1111/0022-4537.00153
Clifford Nass, Jonathan Steuer and Ellen R. Tauber, 'Computers are social actors', in *ACM Proceedings of SIGCHI '94 Human Factors in Computing Systems* (1994), 72–78. https://doi.org/10.1145/259963.260288
[85] Lyle Moran, 'Lawyer cites fake cases generated by ChatGPT in legal brief', *LegalDive,* 30 May 2023. https://tinyurl.com/bdcxxwkj
[86] Emily LaRosa and David Danks, 'Impacts on trust of healthcare AI', in *ACM Proceedings of the 2018 AAAI/ACM Conference on AI, Ethics, and Society*, (2018), 210–215. https://doi.org/10.1145/3278721.3278771
[87] Masahiro Mori, 'The uncanny valley', (1970). The original essay by Masahiro Mori (K. F. MacDorman & N. Kageki, Trans.), IEEE Spectrum. https://spectrum.ieee.org/automaton/robotics/humanoids/the-uncanny-valley

The sidelining of human oversight by AI is detrimental to ensuring justice[88], particularly in sensitive areas such as criminal justice where there is a potential for surveillance and control over individuals[89] or cases relevant to specific areas concerning children's health. [90]Article 14 (4) (b) of the AI Act defines the issues that have raised in this article as automation bias, which is the excessive reliance on algorithmic outputs for high-risk AI systems and further addresses this concern by establishing human oversight mechanism (a similar human oversight mechanism is available also in the GDPR's Article 22 (3)). Even though it is limited to high-risk AI systems, it is a legal requirement and an ethical code to ensure and AI-L is one component of the trust-enhancing principles outlined in the AI Act.

### 5.2. Risk II: Technical Impossibility of Eliminating Bias

From a technical standpoint, eliminating bias and discrimination in legal AI systems involves two major challenges. The first challenge stems from the distinct nature of legal systems. For instance, the civil law tradition evaluates cases individually rather than relying on precedents as in common law systems. This lack of standardized data complicates the training of AI models for civil law systems, making it difficult to achieve accuracy while avoiding bias. As previously discussed elsewhere, determining the desired level of accuracy and the reasons and parties deciding on the accuracy level remain already challenging. Additionally, the inconsistencies within and across national courts and the CJEU impede the standardization necessary for fair AI systems in law[91]. This phenomenon is analogous to the initial unfairness of training data. Furthermore, legal reasoning cannot be confined to mere textual analysis of laws or regulations, as pointed out[92], the content of the applicable law extends beyond the literal meaning of legal provisions as LLMs which rely heavily on textual inputs lack the ability to comprehend the underlying principles driving judicial decisions. LLMs, the current state-of-the-art legal AI model, frequently operate in environments already marked by structural injustices in textual form, further exacerbating existing inequalities within a specific society.[93] This can result in the intensification of existing inequalities rather than their reduction.[94] While numerous biases already exist, only a few can be mitigated, with the rest necessitating technical solution**s.**

While the AI Act introduces provisions for addressing bias, such as requiring providers and deployers to conduct bias evaluations (Article 10) and implement mitigation techniques for high-risk AI systems (Article 15), it has faced notable criticism. Wachter[95] argues that the Act's narrow focus on compliance and technical measures overlooks broader societal and ethical considerations. In areas like gender equality and non-discrimination, measures such as bias audits and fundamental rights impact assessments have been deemed insufficient[96]. These shortcomings complicate organizations' efforts to effectively address bias and design robust AI-L programs. Moreover, the Act's reliance on providers for self-assessment and mitigation risks exacerbating these challenges, potentially undermining its objectives. Such limitations not only hinder the development of comprehensive AI-L programs but also create uncertainty for organizations striving to balance compliance with the need to address the broader societal impacts of legal AI.

---

[88] Sarah Valentin, 'Impoverished algorithms: Misguided governments, flawed technologies, and social control', 46(2), *Fordham Urban Law Journal*, (2019), 364–427. https://ir.lawnet.fordham.edu/ulj/vol46/iss2/4/
[89] Fredrika Björklund, 'Trust and surveillance: An odd couple or a perfect pair?', in Lora Anne Viola and Paweł Laidler (eds.), *Trust and transparency in an age of surveillance,* Routledge, (2021),183–200. https://doi.org/10.4324/9781003120827
[90] Juan David Gutiérrez, 'ChatGPT in Colombian Courts: "Why we need to have a conversation about the digital literacy of the judiciary', *Verfassunblog,* 23 February 2023, https://verfassungsblog.de/colombian-chatgpt/
[91] Sandra Wachter, Brent Mittelstadt and Chris Russell, 'Why fairness cannot be automated: Bridging the gap between EU non-discrimination law and AI', 41, *Computer Law & Security Review,* (2021), 105567. https://doi.org/10.1016/j.clsr.2021.105567
[92] Egidija Tamošiūnienė, Žilvinas Terebeiza and Artur Doržinkevič, 'The possibility of applying artificial intelligence in the delivery of justice by courts', 17(1), *Baltic Journal of Law & Politics*, (2024), 223–237. https://doi.org/10.2478/bjlp-2024-0010
[93] Karen Yeung and Adam Harkens, 'How do 'technical' design-choices made when building algorithmic decision-making tools for criminal justice authorities create constitutional dangers? Part II', (April), *Public Law,* (2023), 448-472. https://tinyurl.com/ycy39trr
[94] Aniket Deroy and Subhankar Maity, 'Questioning biases in case judgment summaries: Legal datasets or large language models?', *arXiv Preprint.* https://doi.org/10.48550/arXiv.2312.00554

### 5.3. Risk III: Absence of Proof Regarding Time and Cost Efficiency

While it has been previously discussed that legal AI has the potential to improve time and cost efficiency, a closer examination reveals significant challenges that undermine these assumptions. Key issues include the high costs of development and implementation, environmental impacts, limited data accessibility, and societal resistance, all of which contribute to inefficiencies in adopting such technologies. The reliance on natural language processing (NLP) technologies in legal applications exacerbates these challenges. As Francesconi[97] observes, the absence of structured and digitized legal data substantially increases the financial burden of creating AI models tailored to legal contexts. This challenge is further exacerbated by the significant cost difference between off-the-shelf AI systems and custom solutions designed for specific judicial environments. General purpose models are becoming increasingly expensive to train, making them financially out of reach for companies with limited budgets.[98] Furthermore, the scarcity of publicly accessible and adequate legal data adds to the complexity, as obtaining relevant datasets often requires significant effort and resources. Many law practices in Europe consist of sole practitioners or small firms that lack the infrastructure to generate meaningful work data or capture metadata from legal documents. The absence of universally accessible tools, coupled with jurisdiction-specific markets across the EU, limits the adoption of consistent case management solutions. This fragmentation, along with insufficient resources and limited training data for NLP systems, hampers the development and scalability of advanced legal technologies. Adding to this burden is the need to ensure cybersecurity, as mitigating risks such as data breaches and cyberattacks demands substantial financial investment[99].

Beyond financial costs, AI systems also pose significant environmental challenges. The computational power required for training large-scale models and operating data centers is highly energy-intensive and frequently relies on non-renewable energy sources, thereby amplifying their environmental footprint[100]. For instance, training advanced models like GPT-4 has been reported to consume considerable amounts of energy and water, driving up utility costs for local communities[101]. While the implementation of better regulatory frameworks could help mitigate these environmental impacts, no comprehensive global standards currently exist. The carbon emissions generated during the training of large-scale models rival those of major industrial activities, initiating the urgent need for sustainable AI practices. Furthermore, the governance and development of legal AI systems often reflect structural inequalities. These systems are often dominated by political elites and technical experts, characterized by high implementation costs and limited public engagement, which undermines inclusivity and transparency.[102] This centralized governance model reinforces existing inequities and impedes the widespread adoption and scalability of legal AI solutions. Finally, there is no consensus among legal professionals on the acceptance of legal AI systems[103], which could ultimately nullify all the costs associated with developing these systems.

---

[95] Sandra Wachter, 'Limitations and loopholes in the EU AI Act and AI liability directives: What this means for the European Union, the United States, and beyond', 26(3), *Yale Journal of Law and Technology*, (2024) 671–718. wachter_26yalejltech671.pdf
[96] Lütz (2024).
[97] Enrico Francesconi, 'The winter, the summer, and the summer dream of artificial intelligence in law', 30, *Artificial Intelligence and Law,* (2022). https://doi.org/10.1007/s10506-022-09309-8
[98] Nestor Maslej et al.,'The AI Index 2024 Annual Report', *AI Index Steering Committee, Institute for Human-Centered AI, Stanford University,* (2024). https://doi.org/10.48550/arXiv.2405.19522
[99] Eddie Segal, 'The impact of AI on cybersecurity', *IEEE Computer Society,* 31 March 2024. https://tinyurl.com/4sya5rh6
[100] Philipp Hacker, 'Sustainable AI regulation', 61(2), *Common Market Law Review,*(2024), 345-386. https://doi.org/10.54648/cola2024025

### 5.4. Persisting Opacity: Technical and Social Dimensions

Regardless of the explainable AI tool that is available and in practice, explainability in technical and social terms will continue to be an area of concern. The data in machine learning systems is currently processed differently at each layer, where layers interact through interconnected neurons or nodes, and data transitions are linearized from nonlinear interactions.[104] The system determining the weight and influence of specific data in real-world applications, providing summarized responses to queries[105], will still translate complex, multidimensional relationships into a language only the machine can interpret, not a human. There is still no traceable list of data and processes in which the system relied on answering to a query or to an input or to a prompt.[106] The complexity further increases when thousands of data points form intricate relationships that are incomprehensible to humans. Models, especially those developed for intricate tasks, increase parameter counts to establish more complex relationships between inputs and outputs, and even simple decision tree models can occasionally produce inexplicable results for similar reasons.[107] Consequently, the black-box problem is real, and the explainability that was previously discussed is constrained by technical limitations.

The social aspect of non-practicable explainability is rooted in its technical opacity as well. The technical complexity mentioned above is the primary reason for the challenges in clear communication with the public.[108] If citizens cannot understand the technical aspects, they may attempt to interpret the provided explanations through their own understanding. However, research suggests that this introduces significant risks to the comprehensibility of AI systems[109] causing more misunderstandings than explanations. Additionally, companies that profit from these systems often invoke trade secrets or competitive advantage claims to avoid disclosing algorithmic details, a challenge extensively discussed in the work of Burrel[110] earlier. The AI Act, despite being the latest legislation, might not fully resolve these issues. Wachter's further analysis[111] highlights that the AI Act's transparency framework is undermined by its failure to address the black-box problem, the prioritization of regulatory documentation over user-centric explanations, and the overextension of trade secret protections that obscure critical algorithmic details. The combination of technical and practical explainability problems persists making it impossible to fully ensure transparency rules.

---

[101] Christopher Harper, 'Using GPT-4 to generate 100 words consumes up to 3 bottles of water — AI data centers also raise power and water bills for nearby residents', *Tom's Hardware,* 19 September 2024. https://tinyurl.com/mr34dx58

[102] Nardine Alnemr, 'Democratic self-government and the algocratic shortcut: The democratic harms in algorithmic governance of society', 23, *Contemporary Political Theory*, (2024), 205–227. https://doi.org/10.1057/s41296-023-00656-y

[103] Richard Susskind, '*Tomorrow's lawyers: An introduction to your future*', Oxford: Oxford University Press, (2013). ISBN: 9780199668069.

[104] Matt Taddy, 'The technological elements of artificial intelligence', in Ajay Agrawal and Joshua Gans and Avi Goldfarb (eds.), *The economics of artificial intelligence: An agenda*, National Bureau of Economic Research-University of Chicago Press, (2019), pp. 61–87.

[105] Ethem Alpaydın, '*Machine learning: The new AI*', MIT Press, (2016). ISBN: 9780262529518.

[106] Matthias (2004)

[107] Hamon et al. (2022)

[108] Ibid.

[109] Michael Chromik et al., 'I think I get your point, AI! The illusion of explanatory depth in explainable AI'. in Proceedings of the 26th International Conference on Intelligent User Interfaces (IUI '21), (2021), pp. 1–10. ACM. https://doi.org/10.1145/3397481.3450644

Finally, the use of LLM in courts further exacerbates the explainability problem, such as determining whether the evidence is AI-generated and whether it is hallucinated, thereby undermining the neutrality of LLMs from the perspectives of model neutrality and data neutrality. For legal AI systems, it is insufficient to merely extract insights from pre-trained models that indirectly explain the model's performance based on metrics. Legal AI systems must be explainable in accordance with the rule of law, not persuasively in a factual sense by referring to relevant laws and precedents. The Court of Justice's *Schufa* decision[112] highlighted concerns highlighting the need for explainability in such models, ensuring that individuals can understand how decisions are made and have the opportunity to contest inaccurate or discriminatory assessments. In short, explanations to be provided in legal AI systems should make legal sense for professionals, organizations, and individuals, but this is not the current state of practice.

## 6. Conclusion and Discussion

This article explored the concept of AI-Literacy as referred to in the Article 4 of the EU AI Act, the nature of legal AI systems, and their intersection through a risk and benefit analysis which is a critical requirement also in essence of the AI Act. While the potential benefits of legal AI systems include the elimination of bias and discrimination, enhanced time and cost efficiency, and improved explainability, these same areas also feature prominently in risk assessments. This dual nature reveals the paradoxical challenges inherent in legal AI systems, which present opportunities to address longstanding problems while simultaneously introducing new risks.

This inherent complexity, in combination with legal obligations, reflects the necessity of equipping legal professionals and developers with AI-L at all stages of a system's life cycle, from design to market placement and deployment. This approach adopts AI-L is not merely a technical or legal requirement but an integral part of organizational culture and societal understanding, encompassing ethical considerations and risk assessments. The findings of this article demonstrate that effective communication is paramount to AI-L, facilitating stakeholder engagement, fostering trust, and enabling informed decision-making. The employment of targeted questionnaires, for example, could be a tool for fostering a participatory environment where transparency and collaboration can flourish.

To operationalize these concepts, the annexed questionnaire, titled "Legal AI Literacy Program for Organizations: A Roadmap Questionnaire," offers a pragmatic framework for initiating AI-

---

[110] Jenna Burrell, 'How the machine 'thinks': Understanding opacity in machine learning algorithms', 3(1), *Big Data & Society*, (2016). https://doi.org/10.1177/2053951715622512
[111] Wachter (2024)
[112] Judgment of 7 December 2023, SCHUFA Holding, EU:C:2023:957.

L initiatives. By promoting stakeholder engagement and customizing literacy programs to the distinct requirements of organizations, the questionnaire offers actionable insights. The questionnaire's hypotheses can be evaluated during system development by providers or prior to implementation by deployers, ensuring that organizational practices align with ethical and regulatory expectations.

The insights and the questionnaire presented in this article aim to pave the way toward a more informed and equitable future. The integration of AI-L at all levels of an organization, and subsequently in society, is essential to harnessing the benefits of legal AI systems while mitigating their risks, creating a balanced approach that serves both organizational efficiency and the broader public interest. With this in mind, the ugly truth, as operationalized through the risk assessment presented in this article, could turn out to be good and beneficial before eventually becoming bad for the organizations.

**Annex:**

**Legal AI Literacy program for the organizations: A roadmap questionnaire**

General questions at the initial phase of the AI-L training module:
1. What are the legal and ethical frameworks that apply to the legal AI system?
2. What are the human rights implications of developing and deploying legal AI systems within organizations?

**AI-L Component I: Bias and Discrimination**

1. What are bias and discrimination in AI systems, particularly in legal AI systems, and how do they manifest?
2. What are the possible scenarios in which legal AI systems might generate biased outputs, and what measures are in place to mitigate them?
3. What auditing tools are available to detect bias and discrimination in legal AI systems, and how do they operate?
4. Does the legal AI system contribute to eliminating bias and discrimination, or could it potentially exacerbate these issues?
5. How can organizations ensure that training data for AI systems is representative and free of systemic biases?
6. To what extent does the use of AI in organizations improve consistency and fairness compared to human decision-makers, and how can organizations provide evidence of these improvements?
7. What policies and/or strategies exist for rejecting legal AI system outputs suspected of being biased, and how are such suspicions handled?
8. What strategies do the organizations adopt to mitigate overreliance on legal AI systems?
9. What tools and methods are available for educating legal professionals about monitoring and refining AI systems to mitigate bias over time?
10. If the legal AI system operates with LLMs, how does it ensure compliance with the jurisdiction of operation?
11. What cross-disciplinary approaches can organizations adopt to integrate technical, legal, and ethical perspectives? How can these approaches help address bias and discrimination in legal AI systems?

**AI-L Component II: Time-cost efficiency**

1. Would it cost less to deploy a human evaluator, or is a human-legal AI collaboration more time and cost-efficient?
2. Are there reasons beyond cost and efficiency, such as better alignment with public interests, for choosing one approach over the other?
3. How can organizations ensure that cost savings achieved through AI implementation align with fundamental rights?
4. Why is this legal AI system more cost-efficient than relying on humans?
5. In which areas is the time and cost efficiency of the legal AI system least risky to human rights, making it suitable for deployment?
6. What strategies are available if deployers or individuals prefer not to use the legal AI system?
7. How can organizations balance cost and time efficiency with producing accurate and reliable outputs?
8. Does the legal AI system contribute to the time and cost efficiency of the organization?
9. Is the legal AI system environmentally friendly? What factors were considered during its development to ensure environmental sustainability, and how can these be maintained during its use?
10. Why was this specific model chosen over alternatives from the perspectives of time, cost, and environmental considerations?
11. Has multilingualism been taken into account in the operation of the legal AI system?
12. What are the potential concerns of legal professionals regarding the use of the legal AI system, and how can these be mitigated?
13. Was a legal expert involved in overseeing the legal AI system during the testing phase?

**AI-L Component III: Explainability**

1. In what ways does the legal AI system improve explainability for human professionals?
2. What explainability methods and tools were followed and why?
3. Can the legal AI system dynamically adjust its explainability to the needs of the user, or does it generate a general explanation?
4. What frameworks and methodologies did the organizations adopt to ensure a human-centered approach to designing explainable AI systems for legal applications, or does it provide only what the model can generate?
5. Is it possible for users to request personalized explanations?
6. How can explainable AI techniques be adapted to meet the diverse expertise levels of different users, from legal experts to laypersons?
7. How does the legal AI system ensure legal professionals can effectively collaborate with AI systems while maintaining their creative and interpretative roles?
8. Can the legal AI system offer personalized explanations while maintaining accuracy and efficiency? Why the model is assigned the specific accuracy metrics by the organization?
9. What educational resources are available to help legal professionals effectively use and critically evaluate explainable AI tools in their work?

**Acknowledgment**

The research was supported by the ICT and Societal Challenges Competence Centre of the Humanities and Social Sciences Cluster of the Centre of Excellence for Interdisciplinary Research, Development and Innovation of the University of Szeged.

**Re, R. M., & Solow-Niederman, A.** (2019). Developing artificially intelligent justice. *Stanford Technology Law Review, 22*(2), 242–289.

**Rebstadt, J., Remark, F.,Fukas, P., Meier, P., and Thomas, O.,** (2022). Towards personalized explanations for AI systems: Designing a role model for explainable AI in auditing. *Wirtschaftsinformatik 2022 Proceedings, 2*. Retrieved from https://aisel.aisnet.org/wi2022/ai/ai/2

**Reed, C., Grieman, K., & Early, J.** (2022). Non-Asimov explanations: Regulating AI through transparency. *The Swedish Law and Informatics Research Institute, 1*, 315–338. https://doi.org/10.53292/208f5901.20b0a4e7

**Saunders, T.** (2023, October 19). Legal tech teams turn to AI to advance business goals. *Financial Times*. Retrieved from https://www.ft.com/content/9a117ac7-29ae-43fe-b840-a04005b98799

**Segal, E.** (2024). The impact of AI on cybersecurity. *IEEE*. Retrieved from https://www.computer.org/publications/tech-news/trends/the-impact-of-ai-on-cybersecurity

**Shi, C., Sourdin, T., & Li, B.** (2021). The smart court – A new pathway to justice in China? *International Journal for Court Administration, 12*(1), 4. https://doi.org/10.36745/ijca.367

**Shneiderman, B.** (2020). Bridging the gap between ethics and practice: Guidelines for reliable, safe, and trustworthy human-centered AI systems. *ACM Transactions on Interactive Intelligent Systems, 10*(4). https://doi.org/10.1145/3419764

**Sourdin, T.** (2021). *Judges, technology, and artificial intelligence*. Elgar Law, Technology, and Society. eISBN: 9781788978262.

States v. Loomis (881 N.W.2d 749 (Wis. 2016)

**Steenhuis, Q., & Westermann, H.** (2024). Getting in the door: Streamlining intake in civil legal services with large language models. *arXiv Preprint*. Retrieved from https://arxiv.org/abs/2410.03762

**Susskind, R.** (2013). *Tomorrow's lawyers: An introduction to your future*. Oxford Scholarship Online.

**Susskind, R.** (2019). *Online Courts and the Future of Justice* (New York, 2019; online edn, Oxford Academic), https://doi.org/10.1093/oso/9780198838364.001.0001

**Taddy, M.** (2019). The technological elements of artificial intelligence. In A. Agrawal, J. Gans, & A. Goldfarb (Eds.), *The economics of artificial intelligence: An agenda* (pp. 61–87). University of Chicago Press.

**Tamošiūnienė, E., Terebeiza, Ž., & Doržinkevič, A.** (2024). The possibility of applying artificial intelligence in the delivery of justice by courts. *Baltic Journal of Law & Politics, 17*(1), 223–237. https://doi.org/10.2478/bjlp-2024-001028

https://uk.westlaw.com/Document/I3AF53F4004CA11EEAEE6B640426E812F/View/FullText.html